# An Efficient Optimal-Equilibrium Algorithm for Two-player Game Trees

**Michael L. Littman**
Dept. of Computer Science
Rutgers University
Piscataway, NJ 08854

**Nishkam Ravi**
Dept. of Computer Science
Rutgers University
Piscataway, NJ 08854

**Arjun Talwar**
Dept. of Mathematics
Stanford University
Stanford, CA 94309

**Martin Zinkevich**
Dept. of Computing Science
University of Alberta
Edmonton Alberta
Canada T6G 2E8

## Abstract

Two-player complete-information game trees are perhaps the simplest possible setting for studying general-sum games and the computational problem of finding equilibria. These games admit a simple bottom-up algorithm for finding subgame perfect Nash equilibria efficiently. However, such an algorithm can fail to identify optimal equilibria, such as those that maximize social welfare. The reason is that, counterintuitively, probabilistic action choices are sometimes needed to achieve maximum payoffs. We provide a novel polynomial-time algorithm for this problem that explicitly reasons about stochastic decisions and demonstrate its use in an example card game.

## 1 Introduction

Game-tree search for zero-sum games has been a staple of AI research since its earliest days. Recently, research on general-sum games has intensified as a way of reasoning about more complex agent interactions (Kearns et al. 2001). In this paper, we treat the problem of finding optimal Nash equilibria in general-sum two-player game-trees.

### 1.1 Problem Definition

A game-tree problem, as studied in this paper, is specified by a tree with $n$ nodes. A leaf node has no children, but it does have a payoff vector $R(i) \in \Re^2$. Each non-leaf node $i$ has an associated player $T(i) \in \{1, 2\}$ who controls play in that node. Except for the root node, each node has a parent node $P(i) \in \{1, \ldots, n\}$.

Play begins at the root of the tree. When playing node $i$, Player $T(i)$ chooses an action. We define the set of actions available from node $i$ as $A(i) = \{j | P(j) = i\}$; Player $T(i)$ selects from among the child nodes of $i$. When a leaf node $i$ is reached, the players receive their payoffs in the form of a payoff vector $R(i)$. Specifically, Player $x$ receives the $x$th component of the payoff vector $R(i)_x$.

In the small example in Figure 1, node numbers are written above the nodes and leaves are marked with rectangles. Non-leaf nodes contain the player number for the player who controls that node and leaf nodes contain their payoff vectors. Here, play begins with a decision by Player 2 at node 1. Player 2 can choose between the two children of the root, node 2 and node 3. If Player 2 selects node 2 as his action, Player 1 gets to make a choice between node 4 and node 5. If Player 1 chooses the left child, node 4, a leaf is reached. From node 4, Player 1 receives a payoff of 2 and Player 2 receives a payoff of 3 and the game ends.

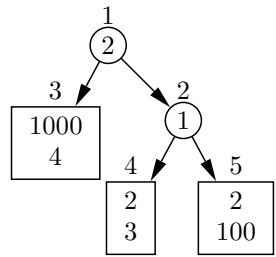

Figure 1: A small game tree that demonstrates the challenge of finding optimal equilibria.

### 1.2 Equilibria

A *joint strategy* for a game tree is a selection, at each node of the game tree, of a probability distribution over actions. We can write a joint strategy as a function $\pi$ that maps each node of the tree to a probability distribution over its children. Returning to Figure 1, one joint strategy is $\pi(1)_2 = 1/2$, $\pi(1)_3 = 1/2$ (choose node 2 from node 1 with probability 1/2 and node 3 with probability 1/2), and $\pi(2)_5 = 1$ (always choose node 5 from node 2).

The *value* of a strategy $\pi$ is the expected payoff vector if strategy $\pi$ is followed. It can be defined recursively as follows. In the context of a strategy $\pi$, define $V^\pi(i)$ to be the value for the game defined by the subtree rooted at node $i$. For a leaf node, $V^\pi(i) = R(i)$. Otherwise,

$$V^\pi(i) = \sum_{j \in A(i)} \pi(i)_j V^\pi(j). \qquad (1)$$

The sum, here, is taken over each component of the payoff vector separately. Equation 1 just tells us that once we calculate the values for each child node, we can find the values of each parent and consequently the root by working our way up the tree.

The example strategy above has a value of $[501, 52]$ because, starting from node 2, Player 1 chooses node 5 resulting in payoffs of $[2, 100]$. Starting from node 1, Player 2 randomizes over node 2 and node 3, resulting in the expected payoffs given.

A joint strategy $\pi$ is a *subgame perfect Nash equilibrium* (Osborne and Rubinstein 1994), or *equilibrium* for short, if there is no $i$ such that $T(i)$ can increase her value $V^\pi(i)_{T(i)}$ at $i$ by changing the strategy $\pi(i)$ at $i$. Specifically, let $\pi$ be a strategy for a game, and $i$ be a node of the game tree. We say that strategy $\pi$ is *locally optimal* at node $i$ if

$$V^\pi(i)_{T(i)} \geq \max_{j \in A(j)} V^\pi(j)_{T(j)}. \qquad (2)$$

Equation 2 simply states that the player controlling node $i$ ($T(i)$) can't improve her payoff by selecting some other action choice at node $i$. Strategy $\pi$ is an equilibrium if it is locally optimal at every node.

A *subgame* at node $i$ of a game tree is the game obtained by retaining only the subtree rooted at node $i$. An equilibrium strategy for a game is also an equilibrium strategy for all subgames of the game.

Note that the example strategy given above is *not* an equilibrium because Player 2 can increase his payoff to 100 by choosing node 2 deterministically from node 1. The joint strategy $\pi_A(1)_2 = 1$, $\pi_A(2)_5 = 1$, is an equilibrium, however, as Player 1 cannot improve her value of 2 by changing her decision at node 2 and Player 2 cannot improve his value of 100 by changing his decision at node 1.

The definition of an equilibrium can be turned directly into an algorithm for computing an equilibrium. Define $V(i) = R(i)$ for every leaf node $i$. For non-leaf nodes, $V(i) = V(j^*)$ where $i = P(j^*)$ and $j^* \in A(i)$ is chosen such that $V(j^*)_{T(i)} \geq \max_{j \in A(j)} V(j)_{T(i)}$. In words, $V(i)$ is the value vector of the child of node $i$ that has the largest payoff for the player controlling node $i$ ($T(i)$). Such a player is happy with its choice at that node since no other choice will improve its expected payoff.

Given the $V$ function, the corresponding equilibrium strategy is $\pi(i)_j = 1$ where $j \in A(j)$ and $V(j) = V(i)$; at each node, the strategy chooses the child whose value vector matches that of the node. The anyNash algorithm returns this strategy, $\pi$.

By construction, the strategy produced by anyNash satisfies the conditions for being an equilibrium expressed in Equation 2. Specifically, at each node $i$ the strategy chooses a child node that results in the best possible (local) outcome for the player who controls node $i$.

### 1.3 Optimal Equilibria

Some games, like the example game in Figure 1, have multiple equilibria. The situation arises initially when a player has a choice of actions that results in identical payoffs for that player. In fact, if anyNash does not encounter any ties as it proceeds up the game tree, it finds the unique equilibrium.

When there is a choice of equilibria, anyNash chooses one arbitrarily. However, some equilibria are more desirable than others. We use the term *optimal equilibrium* to refer to an equilibrium that is the most desirable among the possible equilibria.

There are many possible kinds of desirable equilibria and we focus on four representative types here. If $V_x$ is the value to Player $x$ for playing the equilibrium, we can seek the equilibrium that is the:

1. **social optimum**; maximize the sum of values over all players: $\sum_x V_x$.

2. **fairest**; maximize the minimum payoff: $\min_x V_x$.

3. **single maximum**; maximize the maximum payoff over all players: $\max_x V_x$.

4. **best for** $y$; maximize the payoff to some specific Player $y$: $V_y$.

Greenwald and Hall (2003) refer to these concepts as utilitarian, egalitarian, republican, and libertarian, respectively. They are also explicitly treated by Conitzer and Sandholm (2003) in the bimatrix setting.

The previous section described an equilibrium $\pi_A$ for the example game with a value of $[2, 100]$. A second equilibrium is $\pi_B(1)_3 = 1$ and $\pi_B(2)_4 = 1$. The value of this equilibrium is $[1000, 4]$. Equilibrium $B$ is the social optimum ($1004 > 102$), the fairest ($4 > 2$), the single maximum ($1000 > 100$), and the best for Player 1 ($1000 > 2$). Equilibrium $A$, however, is the best for Player 2 ($100 > 4$).

Note that this game has an infinitely large set of equilibria. In particular, consider a strategy $\pi$ such that $\pi(1)_2 = 1$, $\pi(2)_4 = \alpha$, and $\pi(2)_5 = 1 - \alpha$ (for $0 \leq \alpha \leq 1$). Since $T(2) = 1$ and Player 1 is indifferent to the outcomes at the children of node 2, the subgame rooted at node 2 is an equilibrium, regardless of the value of $\alpha$. For the strategy to be an equilibrium game at node 1, however, it must be the case that $V^\pi(2)_2 \geq V^\pi(3)_2$. Since node 3 is a leaf, $V^\pi(3)_2 = R(3)_2 = 4$. By Equation 1, $V^\pi(2)_2 = \alpha V^\pi(4)_2 + (1-\alpha)V^\pi(5)_2 = \alpha 3 + (1-\alpha)100 = -97\alpha + 100$. We have $V^\pi(2)_2 \geq V^\pi(3)_2$ when $-97\alpha + 100 \geq 4$ or when $0 \leq \alpha \leq 96/97$. Any such value of $\alpha$ leads to a distinct equilibrium strategy.

Later, we show that such stochastic equilibria play a key role in the search for optimal equilibria.

### 1.4 Challenges

When ties are encountered in anyNash, a naive approach to finding an optimal equilibrium is to break ties in favor of the optimal (local) outcome. We note that this approach does *not* generally produce optimal equilibria. As a concrete example, consider trying to find the social optimum in the example game from Figure 1.

anyNash begins with node 2 (the lowest non-leaf node), which Player 1 controls. It is immediately faced with the choice of two actions with equal payoffs for Player 1. Since we seek the social optimum, a natural choice for node 2 is node 5 (total payoffs $102 > 5$). However, now Player 2 is faced with the choice at node 1 of a payoff of 100 for itself or a payoff of 4 for itself. To produce an equilibrium, Player 2 must choose node 2 in this situation, resulting in a total value of 102.

Player 1's selection of node 5 prevented the computation from arriving at the actual social optimum equilibrium, specifically Equilibrium $\pi_B$. This equilibrium would have arisen only from Player 1 selecting node 4. We conclude that an algorithm that wishes to compute the social optimum cannot work by simply selecting actions bottom up—the right choice depends on decisions made elsewhere in the tree. Our algorithm handles this issue by keeping *all* equilibria as it works its way up the tree.

### 1.5 Related Models

We briefly relate the problem we address in this paper to others attacked in the literature.

First, a more general game tree can also have *stochastic nodes* that result in a transition to a child node according to a given set of probabilities. A game without stochastic nodes can be called *deterministic*. Such nodes bridge the gap to richer models such as stochastic games (Shapley 1953, Condon 1992).

A general game tree can also include *information sets*. An information set is a set of nodes controlled by the same player that the player cannot distinguish between. Therefore, the player is constrained to choose the same action or probability distribution over actions at every node in an information set. A game tree without information sets can be called a *complete-information* game.

A game is called *zero sum* if it has two players and, for each leaf node $i$, $R(i)_1 + R(i)_2 = 0$. Other games can be called *general sum*.

Another natural extension is to consider games with more than 2 players.

Equilibria in zero-sum game trees with stochastic nodes and information sets can be found in polynomial time using a linear-programming-based algorithm due to Koller et al. (1996).

General-sum game trees with information sets include bimatrix games as a special case; the root of the tree is the first player's action, the level below is the second player's action, and all nodes at the second level are in the same information set. It is well known that finding optimal equilibria in these games is NP-hard (Gilboa and Zemel 1989, Conitzer and Sandholm 2003). It was also recently shown that finding *any* equilibria in these games is PPAD-hard (Chen and Deng 2005) and is therefore presumed intractable.

A single arbitrary equilibrium can be found in general-sum complete-information game trees with stochastic nodes in polynomial time using a variation of the anyNash algorithm described above. We have shown that finding optimal equilibria, for example social-optimal equilibria, is NP-hard; see the following section.

The problem of finding optimal equilibria in general-sum deterministic complete-information game trees, we show in this paper to be efficiently solvable. Huang and Sycara (2003) referred to this class of games as *complete-information extensive games* (CEGs) and they developed a learning algorithm based on anyNash. This bottom-up approach cannot find optimal equilibria for game trees requiring stochastic actions, such as the example we describe in Section 2.

Our efficient algorithm is specific to two-player games. The problem of finding optimal equilibria in general-sum deterministic complete-information game trees with three or more players is currently an open problem. Figure 2 summarizes the known complexity results for finding equilibria in game trees.

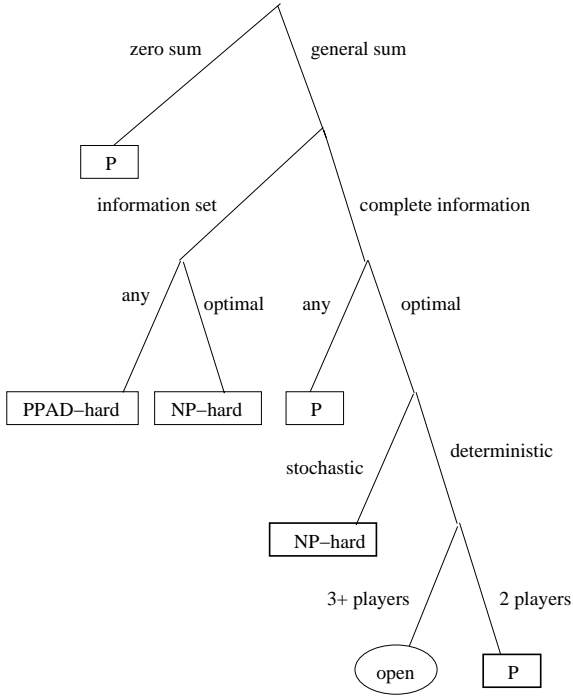

Figure 2: A summary of complexity results for game trees. The heavy boxes are contributions of the current paper.

### 1.6 NP-hardness of game trees with stochastic nodes

Space precludes a formal proof, but we quickly sketch the result.

**The knapsack problem**: There is a given knapsack (Garey and Johnson 1979) of some capacity $c > 0$ and a finite list of $n$ objects, that is, pairs $(w_i, v_i)$ where $w_i$ represents the weight of the object and $v_i$ the value. Find the selection of items ($\delta(i) = 1$ if selected, else 0) such that $\sum_{i=1}^{n} \delta(i) w_i \leq c$ and $\sum_{i=1}^{n} \delta(i) v_i$ is maximized.

Given a knapsack instance, we construct the following game. Player 1 first decides to commence the game or abort in which case the payoffs are $\frac{-c}{nM}$ for Player 1, for large enough $M$ such that the utility of Player 1 is not significant in determining the socially optimal equilibrium, and a very large sum for Player 2. If the game commences, Player 2 is sent to a node of the game tree representing one of the $n$ objects at random (with equal probability), and chooses whether to quit with $[0,0]$ or pass to Player 1 who decides if this object goes in the knapsack or not. Both these actions cost Player 1 $\frac{-w_i}{M}$. However, if Player 1 assigns an object to the knapsack, Player 2 earns $v_i$, and zero otherwise (see Figure 3).

In this game, the subgames initiated by Player 2 for each object have values $[0, 0]$ and $[\frac{-w_i}{M}, v_i]$ as outcomes of their local equilibria. There are several other equilibria as well, but we see that the others cannot improve these payoffs for any player. That is, these are the pareto-dominant equilibria. This property will allow us to equate pareto-dominant equilibria of the whole game to the $2^n$ assignments of the knapsack. Consequently, because of the initial move of the game, the socially optimal solution to the game is equivalent to a knapsack solution. So, a polynomial solution to the stochastic game-tree problem could be used to solve the NP-hard knapsack problem; it is also NP-hard.

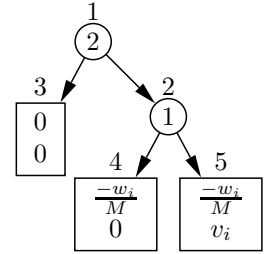

Figure 3: The subgame representing an object with weight $w_i$ and value $v_i$. Any equilibrium where Player 2 gets a strictly positive utility, Player 1 gets a utility of $\frac{-w_i}{M}$.

## 2 The Optimal-Equilibrium Algorithm

The fundamental concept behind our algorithm is that of the *utility profile set* or UPS. The utility profile set for the subgame at node $i$ is

$$U(i) = \{V^\pi(i) | \pi \text{ is an equilibrium strategy}\},$$

the set of possible equilibrium payoffs for the two players. We show that these sets can be represented and manipulated efficiently.

Without loss of generality, we can assume that the game tree is binary: that for each non-leaf node $i$, there are two children, left($i$) and right($i$).[1] Our algorithm, bestNash, for finding optimal equilibria involves first finding the utility profile sets for all subgames.

Algorithm 1 gives the high-level structure of the technique. It computes the UPSs for the children of $i$,

---
[1] If in the true tree there are more than two children (say, $m$ children) for a non-leaf node $i$, then $i$ can be replaced with $m-1$ internal nodes $j_1 \ldots j_{m-1}$ with the same player in control as $i$, where the first node $j_1$ gives a choice of taking the first action or going to the second node, the second node $j_2$ is a choice of taking the second action or going to the third node $j_3$, and so forth, until the last node is a choice between the last two actions.

**Algorithm 1** getEquilibriumUPS(Node $i$)
  **if** isLeaf($i$) **then**
    **return** $\{R(i)\}$
  **end if**
  $S^L \leftarrow$ getEquilibriumUPS(left($i$))
  $S^R \leftarrow$ getEquilibriumUPS(right($i$))
  **return** merge($S^L, S^R, T(i)$)

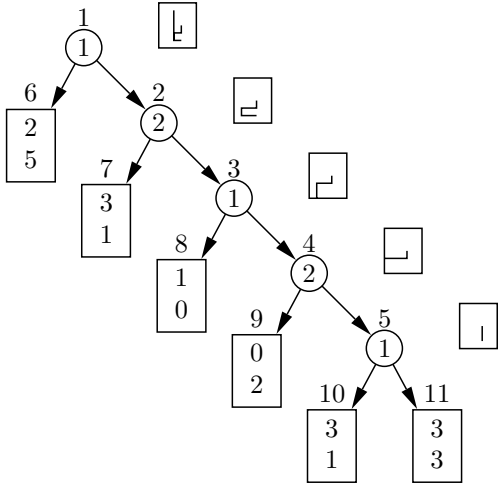

Figure 4: A game where the social optimum equilibrium involves randomness. The square to the right of each node is its UPS.

then combines the results using "merge" to produce the UPS for $i$. Before specifying merge algorithmically, we specify it mathematically:

$$\text{merge}(S^L, S^R, x) = \text{mergeRandom}(S^L, S^R, x) \cup$$
$$\quad \text{mergeLDet}(S^L, S^R, x) \cup \text{mergeLDet}(S^R, S^L, x),$$
$$\text{mergeRandom}(S^L, S^R, x) = \qquad (3)$$
$$\quad \{\lambda s + (1-\lambda)t : \lambda \in [0,1], s \in S^L, t \in S^R, s_x = t_x\},$$
$$\text{mergeLDet}(S^L, S^R, x) = \{s : s \in S^L, s_x \geq \min_{t \in S^R} t_x\}.$$

In words, mergeLDet takes the UPS for one child and keeps only the payoffs that are an improvement for Player $x$ over any payoff in the UPS of the second child. That is, to be an equlibrium payoff, the controlling player must do no less well than the (worst) alternative. However, if Player $x$ has a choice between two equal payoffs, any stochastic combination of the two choices is locally optimal. So, mergeRandom returns any payoff that can be achieved as a convex combination of two payoff vectors for which the controlling player is indifferent. Finally, merge returns any payoff vector that be attained by an equilibrium that deterministically chooses the left child, the right child, or stochastically combines the two.

Consider the behavior of anyNash on the game illustated in Figure 4. From node 5, Player 1 is indifferent and can choose either node 10 or node 11; we consider these two possibilities in turn. If she chooses node 10, the leftward action is preferred by the players at nodes 4, 3, and 2. Ultimately, Player 1 chooses node 2 at node 1, resulting in a payoff vector of $[3, 1]$.

On the other hand, if Player 1 chooses node 11 at node 5, a different cascade takes place. The rightward action is preferred by the players at nodes 4, 3, 2, and 1, resulting in a payoff vector of $[3, 3]$.

On the other hand, if Player 1 randomizes uniformly between nodes 10 and 11 at node 5, $\pi(5)_{10} = \pi(5)_{11} = 1/2$, then $V^\pi(5) = (3, 2)$. Then, if Player 2 randomizes uniformly between nodes 9 and 5 at node 4, $\pi(4)_5 = \pi(4)_9 = 1/2$, then $V^\pi(4) = (1.5, 2)$. The choices at the remaining nodes are deterministic and result in an overall payoff vector of $[5, 2]$. This payoff vector has a higher social welfare than either of the two purely deterministic equilibria, indicating that stochastic action choices must be taken into consideration when searching for optimal equilibria.

Thus, these complex UPSs appear needed to find optimal equilibria. We next explain how to operationalize the mathematical definition given in Equation 4. The key to a practical implementation of merge is that we can *a priori* specify a finite number of regions (based on the payoffs at the leaves), which we refer to as a **basis**, such that any UPS can be represented as a union of some collection of these basis sets. Suppose that $O = \{R(i) | i \text{ is a leaf}\}$ is the set of payoffs. Define $U^1 = \{u_1^1 \ldots u_{n_1}^1\}$ such that $u_1^1 < u_2^1 < \ldots < u_{n_1}^1$ and $U^1 = \{u_1(o)\}_{o \in O}$ to be the list of all $n_1$ possible payoffs for Player 1. Define $U^2 = \{u_1^2 \ldots u_{n_2}^2\}$, $u_2$, and $n_2$ analogously for Player 2. These sets can be constructed in $O(n \log n)$ time, where $n$ is the number of leaf nodes in the tree. Define $N_k = \{1 \ldots k\}$ as the set of numbers from 1 to $k$, for short. Given $U^1$ and $U^2$, define the set of points, lines, and axis-aligned rectangles in the grid defined by $U_1$ and $U_2$ as follows:[2]

$$P_{i,j} = \{(u_i^1, u_j^2)\} \quad \forall i \in N_{n_1}, j \in N_{n_2}, \qquad (4)$$
$$L_{i,j}^1 = [u_i^1, u_{i+1}^1] \times \{u_j^2\} \quad \forall i \in N_{n_1-1}, j \in N_{n_2}, \qquad (5)$$
$$L_{i,j}^2 = \{u_i^1\} \times [u_j^2, u_{j+1}^2] \quad \forall i \in N_{n_1}, j \in N_{n_2-1}, \qquad (6)$$
$$D_{i,j} = [u_i^1, u_{i+1}^1] \times [u_j^2, u_{j+1}^2] \forall i \in N_{n_1-1}, j \in N_{n_2-1}, \qquad (7)$$
$$\mathcal{B} = \{P_{i,j}\} \cup \{L_{i,j}^1\} \cup \{L_{i,j}^2\} \cup \{D_{i,j}\}. \qquad (8)$$

**Definition 1** *A set $S$ is **simply representable** if there exists a collection $\mathcal{B}' \subseteq \mathcal{B}$ such that $S = \cup_{B \in \mathcal{B}'} B$.*

---
[2]Observe that $P_{i,j}, P_{i,j+1} \subseteq L_{i,j}^2$, $P_{i,j}, P_{i+1,j} \subseteq L_{i,j}^1$, and $L_{i,j}^1, L_{i,j+1}^1, L_{i,j}^2, L_{i+1,j}^2 \in D_{i,j}$.

Given a simply representable set $S$, we represent it in the machine by four two-dimensional Boolean arrays, $S.P[\ ][\ ], S.L1[\ ][\ ], S.L2[\ ][\ ]$, and $S.D[\ ][\ ]$, such that:

$$S = \bigcup_{x \in S.P} P_x \cup \bigcup_{x \in S.L1} L_x^1 \cup \bigcup_{x \in S.L2} L_x^2 \cup \bigcup_{x \in S.D} D_x, \tag{9}$$

where the union is over all index pairs for which the two-dimensional array stores the value "true". We prove inductively that the output of getEquilibriumUPS is simply representable. As a base case, for any leaf $o$, $\{(u_1(o), u_2(o))\} = P_{i,j}$ for some $i$ and some $j$. Now, for internal nodes, we need to prove that the output of the merge function, given two simply representable sets, outputs a simply representable set.

**Lemma 2** *If $B, B' \in \mathcal{B}$, then:*

1. *mergeLDet$(B, B', i)$ is simply representable.*
2. *mergeRandom$(B, B', i)$ is simply representable.*

This result can be proven by simple iteration over the various cases.

**Lemma 3** *If $\mathcal{B}^1, \mathcal{B}^2 \subseteq \mathcal{B}$, $S^L = \cup_{B \in \mathcal{B}^1} B$ and $S^R = \cup_{B \in \mathcal{B}^2} B$ (that is, they are simply representable), then merge$(S^L, S^R, i)$ is simply representable.*

This result follows from the distributive property of merge over union, and the fact that the union of two simply representable sets is simply representable.

Thus, we have established that the sets in which we are interested are simply representable. Further, the number of entries in the data structure for representing the set is $\Theta(n_1 n_2) = O(n^2)$.

Applying the distributive property directly to merge simply representable sets is somewhat inefficient, as it takes $\Omega((n_1 n_2)^2)$ time. We can show how it can be done in $O(n_1 n_2)$ time. Algorithm 2 and Algorithm 3 are $O(n_1 n_2)$ algorithms for computing mergeLDet$(\cdot, \cdot, 1)$ and mergeRandom$(\cdot, \cdot, 1)$, respectively at a node for Player 1. Since union can be performed in $O(n_1 n_2)$ time as well, a merge can be performed in $O(n_1 n_2)$ time. Given $U^1$ and $U^2$ (which take $O(n \log n)$ time to construct), one need only perform $n$ merges (one for each node in the game tree), and therefore the overall runtime is $O(n n_1 n_2)$, which is $O(n^3)$, but can be substantially less if the payoffs comes from a small set of possibilities.

To make the merge operation more accessible, let's consider a concrete example. Figure 5 gives example UPSs for $S^L$ and $S^R$. The mergeLDet sets show

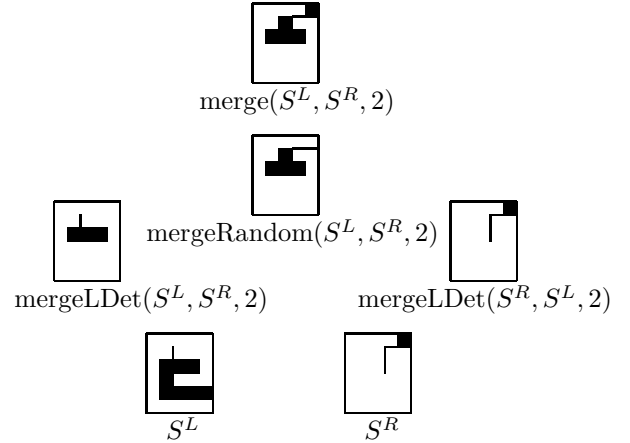

Figure 5: Illustration of the sets generated in the steps of merging two example UPSs.

---

**Algorithm 2** mergeLDetP1(UPS $S^L$, UPS $S^R$)

$m \leftarrow \min\{i : \exists j \text{ s.t. } S^R.P[i][j]\}$
$X \leftarrow \emptyset$
**for** $i = m$ to $n_1$ **do**
  **for** $j = 1$ to $n_2$ **do**
    $X.P[i][j] \leftarrow S^L.P[i][j]$
    **if** $i \neq n_1$ **then**
      $X.L1[i][j] \leftarrow S^L.L1[i][j]$
    **end if**
    **if** $j \neq n_2$ **then**
      $X.L2[i][j] \leftarrow S^L.L2[i][j]$
    **end if**
    **if** $i \neq n_1$ and $j \neq n_2$ **then**
      $X.D[i][j] \leftarrow S^L.D[i][j]$
    **end if**
  **end for**
**end for**

---

the two sets after truncating each one based on the smallest value obtainable in the other subtree. The mergeRandom set comes from "filling in" the points that are obtainable as convex combinations of values for Player 1 when Player 2's values are tied between the two subtrees. Finally, merge is the result of unioning the mergeRandom and mergeLDet sets.

Once the UPS is computed recursively for the root of the game tree, bestNash selects the optimal payoff at the root. For all four notions of optimality considered in this paper, this calculation can be carried out simply by checking the upper-right corner of each of the rectangular regions in the UPS. Other notions of optimality, like maximizing the product of the payoffs of the two players, can also be computed efficiently from the resulting data structure.

**Algorithm 3** mergeRandomP1(UPS $S^L$, UPS $S^R$)

$X \leftarrow \emptyset$
**for** $i = 1$ to $n_1$ **do**
  **if** $\exists j$ s.t. $S^L.P[i][j]$ and $\exists k$ s.t. $S^R.P[i][k]$ **then**
    $low \leftarrow \min\{j : S^L.P[i][j] \text{ or } S^R.P[i][j]\}$
    $high \leftarrow \max\{j : S^L.P[i][j] \text{ or } S^R.P[i][j]\}$
    **for** $j = low$ to $high$ **do**
      $X.P[i][j] \leftarrow \text{true}$
      **if** $j \neq high$ **then**
        $X.L2[i][j] \leftarrow \text{true}$
      **end if**
    **end for**
  **end if**
**end for**
**for** $i = 1$ to $n_1 - 1$ **do**
  **if** $\exists j$ s.t. $S^L.L1[i][j]$ and $\exists k$ s.t. $S^R.L1[i][k]$ **then**
    $low \leftarrow \min\{j : S^L.L1[i][j] \text{ or } S^R.L1[i][j]\}$
    $high \leftarrow \max\{j : S^L.L1[i][j] \text{ or } S^R.L1[i][j]\}$
    **for** $j = low$ to $high$ **do**
      $X.L1[i][j] \leftarrow \text{true}$
      **if** $j \neq high$ **then**
        $X.D[i][j] \leftarrow \text{true}$
      **end if**
    **end for**
  **end if**
**end for**
**return** $X$

Once the optimal payoff is identified, bestNash can proceed recursively top down to create the corresponding equilibrium strategy by choosing the strategy at each node (either a deterministic or stochastic choice as needed) that attains the target value.

## 3 Experimental Results

In this section, we compare the result of applying anyNash to that of bestNash in a game tree derived from a realistic card game.

### 3.1 Rules of the Game

"Oh Hell!" is card game in the Spades family played with closed-handed bidding.[3] Thus, representing the game tree for this game requires information sets.

Since this paper is focused on games of complete information, we modified the rules of the game to create the open-handed variant described below. Note that solutions to complete-information games can be used to guide decision making in imperfect information games (Sturtevant 2004), so this simplification may be of practical interest in computer games.

*Open-Handed Oh Hell! (OHOH)* is played with one standard pack of cards and from 2 to 7 players. On each *hand*, every player is dealt the same number of cards, with the number ranging from 1 to 7. Players can see each other's cards in this variant. Before the game begins, a *trump* suit is randomly chosen. Each player declares the number of *rounds* she thinks she can win; this declaration is called the *contract*. The contracts are declared in a round-robin fashion. A contract can range from zero to the number of cards dealt to a player. The contract of the player who declares last is constrained so that the sum of the contracts of the players does not equal the number of cards dealt—someone will not be able to make her contract.

To win, a player has to meet her contract. That is, the number of rounds won by a player should be the same as her contract. If a player manages to meet her contract, her payoff is 10 plus the value of the contract. If she deviates from the contract, her payoff is $-10$ minus the value of the contract.

A round is played in the following manner: Players play one card each by moving it to the center in a round-robin fashion. The one who goes first is called the *Button* and has the freedom to play any card. The others have to play a card of the same suit, unless they don't have one. If all the cards are of the same suit, the one with the highest value is the winning card. Deuce (2) is the lowest in value and ace is the highest.

If all the cards are not of the same suit, and there is one or more trump cards in the center, the highest trump card is the winning card. If there are no trump cards in the center, and the cards are of different suits, the card that has the color of the Button's card and is highest in value is the winning card. The player with the winning card wins the round, and is designated the Button for the next round.

### 3.2 Empirical Results

We present experimental results on the 2-player, 4-card and 5-card versions of OHOH. We randomly generated 1000 hands. For each hand, we ran anyNash and bestNash.

For each hand we solved, we compared the value of the equilibrium found by anyNash to the one found by bestNash for several notions of optimality. We computed the fraction of times anyNash did *not* find optimal equilibrium. The results, along with other attributes of the games, are summarized in Table 1.

These results show that, even in this simple card game, multiple equilibria abound. Therefore, if we want to find optimal equilibria, we should not rely on anyNash to find them. Although the expense of running bestNash is polynomial, for a large game tree, its worse case running time is substantial (cubic in the number of leaves, where anyNash is linear). However, this example shows that, in practice, the additional expense of running bestNash can be close (within a factor of 6) to that of anyNash.

---
[3] See, for example, http://en.wikipedia.org/wiki/Oh_Hell for rules and other information.

|                              | 4-card        | 5-card        |
|------------------------------|---------------|---------------|
| tree depth                   | 10            | 12            |
| tree size ($n$)              | $\approx 10{,}000$ | $\approx 400{,}000$ |
| distinct payoffs ($n_1, n_2$) | 6            | 7             |
| anyNash running time         | 137 ms        | 2 sec         |
| bestNash running time        | 605 ms        | 11 sec        |
| multiple equilibria          | 30%           | 52%           |
| social optimum               | 22.0%         | 41.5%         |
| fairest                      | 3.2%          | 4.0%          |
| single maximum               | 21.6%         | 39.2%         |
| best for Player 1            | 20.8%         | 37.6%         |
| best for Player 2            | 9.4%          | 17.1%         |

Table 1: Properties of the OHOH game trees and the fraction of the time bestNash found a better equilibrium than anyNash.

Taken together, these results show that bestNash is a viable procedure for searching general-sum game trees.

Another empirical question is whether or not reasoning about action probabilities is critical to finding optimal equilibria. Using the same hands from the experiments above, we implemented a simplified algorithm that finds the optimal *deterministic* equilibrium. This algorithm tended to run about twice as fast as bestNash and half as fast as anyNash. For all 1000 hands, it found the same fairest, single maximum, best for *P1*, and best for *P2* equilibria. However, in 7.6% of the hands for the 4-card game (11.8% for the 5-card game), the deterministic social optimum was worse than that found by bestNash. So, in fact, the probabilistic reasoning did come into play in this natural card game.

## 4 Conclusion

We studied general-sum two-player complete-information game trees and found that it is possible to find optimal equilbria in polynomial time. As shown by a simple example, we discovered that computing optimal equilibria requires reasoning about probabilistic strategies. Our proposed algorithm builds utility profile sets for each node of the game tree from the bottom up and uses the set at the root, which compactly represents all equilibria, to select an optimal equilibrium of any of several types. An open problem is whether an efficient algorithm exists for 3-or-more player games or for other equilibrium concepts such as perfect equilibria or subgame-perfect equilbria that are all socially optimal.

We implemented the new algorithm on a modification of a popular card game and found that, indeed, multiple equilibria are common. Our algorithm was frequently able to find improvements over arbitrarily chosen equilibria and a significant fraction of games required the use of probabilistic reasoning to identify the equilibrium that maximized the total payoffs to the two players.

## Acknowledgments

We thank our anonymous reviewers, Nathan Sturtevant, and the support of NSF ITR IIS-0325281, NSF Research Experience for Undergraduates, and the Alberta Ingenuity Centre for Machine Learning.